# Multi-channel high-speed flip-chip packaging platform for thin-film lithium niobate photonic circuits

Zhenzheng Wang (汪振政)[1,2], Xiangzhi Xie (谢祥芝)[1,2], Yikun Chen (陈逸堃)[1,2], Yifan Wu (吴怡凡)[1,2], Zeyu Luo (罗泽宇)[1,2], Hanke Feng (冯寒珂)[1,2], Yulong Chen (陈宇龙)[1], Zengyi Xu (徐增熠)[3], Zixi Wang (王梓希)[1,2], Yuansong Zeng(曾元松)[1,2], Mandie Huang(黄曼昳)[1], Ke Zhang (张珂)[4], Yingxia Liu (刘影夏)[5], Cheng Wang (王骋)[1,2,]*

[1]*Department of Electrical Engineering, City University of Hong Kong, Kowloon, Hong Kong, China*

[2]*State Key Laboratory of Terahertz and Millimeter Waves, City University of Hong Kong, Kowloon, Hong Kong, China*

[3]*Key Laboratory for the Information Science of Electromagnetic Waves (MoE), College of Future Information Technology, Fudan University, Shanghai, China*

[4]*Rhinopix Technology Limited, Hong Kong, China*

[5]*Department of Systems Engineering, City University of Hong Kong, Kowloon, Hong Kong, China*

**cwang257@cityu.edu.hk*

## ABSTRACT

To address the urgent need for multi-channel high-speed electrical interfacing of thin-film lithium niobate (TFLN) photonic circuits, we realize a flip-chip packaging platform capable of simultaneously delivering 13 high-speed and 32 low-speed electronic signals to a centimeter-sized TFLN chip. The platform exhibits low flip-chip bonding loss and low inter-channel crosstalk over a broad bandwidth up to 50 GHz. Leveraging this packaging platform, we demonstrate high-speed electrical interfacing with two proof-of-concept TFLN photonic circuits, namely a 2×8 optical switch and an electro-optic comb-based transmitter. The switch achieves arbitrary 8-channel routing with ~3 dB insertion loss, < -20 dB crosstalk, and an equipment-limited switching time of ≤ 34 ps. The transmitter circuit includes a 50 GHz electro-optic comb generator with 2.8-dB flatness, a tunable microring to arbitrarily filter one comb line, and a modulator for data transmission at 20 Gbit/s. The packaging platform could significantly advance large-scale TFLN circuits in optical communications, microwave photonics, and photonic computing.

## INTRODUCTION

Thin-film lithium niobate (TFLN) platform has emerged as an ideal candidate for photonics due to the unique combination of excellent material properties, such as ultra-high electro-optic (EO) response speed, wide transparency window (0.4-5.5 μm), low optical loss (~0.001 dB/cm), large diagonal $\chi^{(2)}$ nonlinearity (~30 pm/V), and strong optical confinement in a sub-micron-thick device layer[1-4]. These attributes are generally not simultaneously achievable on other common photonic platforms, such as silicon[5], silicon nitride (SiN)[6], gallium arsenide (GaAs)[7], and polymers[8]. To date, TFLN photonics has enabled a variety of devices with excellent performance, including high-speed EO modulators[9-13], broadband EO frequency comb sources[14-18], high-*Q* microring[19-21] and disk resonators[22], and so on. However, individual standalone devices are insufficient for more advanced system-level applications on chip, such as high-capacity optical communications[23,24], microwave photonics[25,26], and optical computation[27]. Therefore, the move toward larger-scale photonic device integration on the TFLN platform has become an essential trend. Such large-scale circuits necessitate simultaneously interfacing a large number of TFLN optical components with external electronic devices at broad electrical bandwidths beyond 50 GHz.

In academic research environments, external electrical signals are typically fed to TFLN chips via high-speed probes due to testing flexibility and low loss. Limited by the physical sizes of the probe stations and the number of connectors per probe, this approach allows for at most a few high-speed electrical inputs and is often vulnerable to environmental fluctuations[28,29]. As a result, in industrial and system-level testing settings, it is more favorable to package the integrated TFLN chips with multiple input/output (I/O) electrical channels via wire bonding[30,31]. However, high-speed wire-bonding pads are typically required to be placed near the chip edges to reduce the wire length and electrical loss, significantly limiting the allowed number of I/O interfaces. Even so, the electrical loss through the bonding wires could still become increasingly significant at high frequencies[32-34], making it challenging to achieve multi-channel high-speed electrical interfaces. Advanced flip-chip (FC) bonding technologies[32,35-37] could provide an effective solution to this challenge. By directly connecting arrays of electrical pads on two face-to-face chip surfaces through solder balls, this method accommodates substantially more electrical I/Os for a certain chip size. Moreover, direct solder connections could minimize the associated radio frequency (RF) loss and parasitic inductance/capacitance, leading to broad operation bandwidths. It also aligns naturally with 3D-stacked co-packaged optics (CPO)[38,39], a crucial step toward future densely integrated optoelectronic systems.

Here, we demonstrate, for the first time, multi-channel high-speed FC packaging of TFLN photonic integrated circuits. This packaging platform supports 13 high-speed electrical channels for driving EO devices, with low connection loss and inter-channel crosstalk up to 50 GHz, together with 32 low-speed channels for controlling thermo-optic (TO) devices, simultaneously. As proofs of concept, we validate the effectiveness of our FC packaging platform using two distinct TFLN photonic circuits, namely a 2 × 8 optical switch and an EO comb-based optical transmitter. For the optical switch, we show arbitrary switching between the 8 output channels with a low insertion loss ~3 dB, low inter-channel crosstalk < -20 dB, and a switching time no longer than 34 ps, currently limited by the bandwidth of the oscilloscope. For the EO comb-based optical transmitter, we simultaneously achieve the generation of a 50 GHz-spaced EO comb with 2.8-dB power flatness via cascaded amplitude and phase modulators, filtering of one arbitrary comb line using a tunable microring resonator, and high-speed data modulation of the selected comb line at up to 20 Gbit/s (limited by the oscilloscope) using a third EO modulator. These demonstrations establish the proposed FC packaging technology as a reliable multi-channel high-speed electrical interconnect solution for large-scale TFLN photonic circuits, with a broad range of applications including large-scale optical neuromorphic computing, high-capacity optical communications, and ultrafast optical metrology.

## RESULTS AND DISCUSSION

### Overall scheme of the FC packaging platform

Fig. 1a illustrates a conceptual schematic of an envisioned FC-bonded CPO platform, which electrically interfaces large-scale TFLN photonic circuits with an application-specific integrated circuit (ASIC) through an intermediate high-speed printed circuit board (PCB). In this configuration, FC bonding is achieved between the photonic chip's top surface and the PCB's bottom surface through solder balls, which are depicted by exaggerated silver spheres above the optical chip in Fig. 1a for better visualization. In principle, these solder balls can be placed at any desired location on the chip, thereby providing flexible, multi-channel, high-speed, and low-loss electrical interfaces for driving and controlling a variety of functional photonic systems, such as photonic neural networks, multi-channel optical switches, and EO comb-based transmitters. By positioning the ASIC closer to the photonic chip, this configuration not only significantly reduces the electrical propagation loss on the PCB traces, but also potentially allows package-level co-design of the full driving circuit from ASIC to device loads. In addition to the electrical packaging, optical packaging can be further implemented through interconnection between spot-size-converted TFLN waveguides and fiber arrays.

In this work, we take the first critical step toward the envisioned full FC-bonded CPO platform, i.e., FC bonding between the photonic circuit and the PCB, as shown in Fig. 1b, while the ASIC co-packaging and optical packaging will be subject to future research and development.

## Design and electrical characterization of the FC bonds

Fig. 2a shows a back-side picture of the full PCB assembly (PCBA) module, which consists of a PCB and a TFLN photonic chip, FC bonded to each other. At the center of the PCBA lies the bonding region, with magnified views of the corresponding layouts on the PCB and the TFLN chip prior to FC bonding presented in Fig. 2b and Fig. 2c, respectively. As illustrated in Fig. 2b, the PCB incorporates a total of 81 bonding pads, each 250 μm in size with a minimum center-to-center pitch of 500 μm. Among these, 39 pads are arranged along the upper and lower edges of the bonding region, collectively forming 13 channels of high-speed ground-signal-ground (GSG) pads dedicated to fast EO devices. These high-speed pads are fanned out through optimized electrical traces—each achieving an electrical 3 dB bandwidth exceeding 50 GHz—toward two 8-channel high-speed electrical connector arrays, which could be further interfaced with external coaxial cable bundles. The remaining 42 pads are positioned along the left and right edges, providing direct-current (DC) control of on-chip TO heaters. By sharing ground among several signal lines, the 42 bonding pads allow the simultaneous delivery of 32 low-speed electronic signals. These low-speed pads are connected through vias to the back side of the PCB and then routed to alignment holes located at the four corners of the PCB, which are designed to accommodate 10-pin male headers for the low-speed rainbow flat cables. The present channel count is primarily limited by the footprint of the cable connectors. In future iterations, direct mounting of ASICs onto the PCB, as envisioned in Fig. 1a, could lead to even higher channel density and lower interconnection loss. On the photonic side (Fig. 2c), the chip features a dimension of 1.5 cm × 1.5 cm, slightly wider than the PCB bonding area, thereby allowing the input and output optical ports at the chip edges to remain visible from the top side for optical coupling and alignment. The bonding pads on the photonic chip are precisely aligned with those on the PCB in both size and position. The high-speed pads are connected to impedance-engineered GSG transmission lines feeding amplitude modulators (AMs) or phase modulators (PMs), each approximately 1.1 cm in length. At the opposite end, each transmission line is terminated with a 50 Ω on-chip load to suppress electrical reflections. The two demonstrated photonic circuits—a 2×8 optical switch and an EO comb-based transmitter—are located on the left and right sides of the chip, respectively.

Fig. 2d displays a zoom-in cross-sectional schematic of the bonding region. The initial fabrication process of the photonic chip followed that used in our previous work[40], resulting in half-etched 500-nm-thick lithium niobate waveguides, 1-μm-thick Cu/Au transmission line electrodes, nichrome (NiCr) heaters and on-chip loads, and another layer of 1-μm-thick Cu/Au for on-chip routing traces. Next, the entire chip was coated with a 3-μm-thick $SiO_2$ layer. Vias were then opened to expose the bonding pad surfaces, followed by deposition of an additional 2 μm of Cu/Au to thicken the bonding pads. Finally, solder balls were used to attach the chip to the PCB after precise alignment, preheating and vacuum reflow soldering, achieving reliable electrical interconnection. A detailed description of the fabrication and bonding process is provided in the Supplementary data. Fig. 2e presents an X-ray image of the PCBA module after FC bonding, revealing excellent alignment between the PCB and the TFLN photonic chip.

To evaluate the yield of our FC bonding process, a batch of four photonic chips was packaged, incorporating a total of 52 high-speed channels and 52 low-speed channels (the remaining low-speed channels are unused in the current design). Electrical characterization, including EE S11 measurements and DC resistance testing, revealed that all 52 high-speed channels and 46 of the 52 low-speed channels achieved successful bonding, corresponding to an overall bonding yield of 94%. Unsuccessful bonds primarily result from out-of-plane deformation in the central waist region of the PCB, which is narrower than other regions and possibly undergoes non-uniform thermal gradients and stress during the bonding process. These issues could be mitigated in the future by incorporating supporting frames to the narrow bonding area of the PCB, reducing the size of the PCB in a full CPO system, and/or optimizing the thermal sequence during bonding.

We analyzed the electrical loss introduced by FC bonding through a comparative analysis of the EO S21 responses of a TFLN AM before and after packaging. As shown in Fig. 2f, the light gray curve represents the normalized EO S21 of the AM measured at the chip level before packaging using a high-speed probe, exhibiting a 3-dB EO bandwidth of approximately 30 GHz. The EO bandwidth of the present device is partly limited by taper-induced impedance mismatch in the GSG electrodes and imperfections in the fabrication of the Cu/Au electrode layer. After FC bonding, the normalized EO S21 response measured on the packaged PCBA shows slightly increased loss while maintaining a similar roll-off trend (blue). This package-level EO S21 response corresponds to a combination of the chip-level EO S21 response, the electrical loss of the PCB trace (with known values), and the electrical loss of our FC solder joint. After de-embedding the known transmission loss of the PCB trace, we obtain the light blue trace in Fig. 2f that corresponds to the chip-level EO S21 plus the FC solder joint loss. Subsequently subtracting the previously obtained chip-level EO S21 yields the electrical loss (EE S21) exclusively due to the FC solder joints (light red). The raw EE S21 curve is relatively noisy, with minor dips likely arising from parasitic resonances in the FC bonds and/or the electrodes, which can be suppressed through further design and process optimization. To better visualize the underlying trend of the frequency response, we further applied smoothing to the raw data (red dashed curve). Our results suggest that the overall bonding-induced loss remains low across the entire measured frequency range, with an average loss below 1 dB from DC to 53 GHz. EE S11 measurement of the PCBA (gray) shows several reflection peaks across the spectrum, which correlate with the dips observed in the EO S21 curve of the PCBA. This suggests that there is still room for improvement in impedance matching and the FC bonding process.

To assess the electrical crosstalk between channels in the packaged PCBA, we measured the electrical S21 transmission responses between an arbitrarily chosen channel (CH 6) and other channels. These include channels located on the same side (CH 1, CH 4, and CH 5, with lateral distances of 9 mm, 3.6 mm, and 1.8 mm from CH 6, respectively) and on the opposite side (CH 9, CH 10, and CH 13, with a longitudinal distance of 1.1 cm and lateral distances of 0.9 mm, 2.7 mm, and 8.1 mm from CH 6, respectively). All measured crosstalk values within the 0–53 GHz frequency range were below -22 dB, with a consistently decreasing trend for channels with increasing physical separation from CH 6. The minimal inter-channel crosstalk is crucial for applications that require simultaneously driving multiple high-speed EO devices in large-scale photonic circuits.

## Proof-of-concept photonic circuit demonstrations

To verify the practical applicability of the proposed FC bonding platform, we next demonstrate the electrical interfacing of two proof-of-concept TFLN photonic circuits, including a high-speed 2×8 optical switch and an EO comb-based photonic transmitter.

### High-speed 2×8 optical switch

Multi-channel high-speed optical switches capable of rapidly reconfiguring connections among multiple I/O optical channels represent a cornerstone technology for advanced optical communication systems and data center interconnects[41-43]. With the escalating demand for complex and fast reconfigurable optical networks, simultaneously achieving a large number of I/O channels and high-speed switching remains a significant challenge.

Here, we take advantage of the high-speed EO response of TFLN devices and the multi-channel interfacing capability of our FC bonding platform to demonstrate an integrated high-speed 2×8 optical switch, as illustrated in Fig. 3a. The switch consists of three cascaded stages of 7 identical and balanced 2×2 AM units. Each 2×2 AM unit is integrated with both EO and TO tuning electrodes for fast switching and bias controlling purposes, respectively (Fig. 3b). The use of TO bias control could mitigate the well-known electro-optic DC drift issue in lithium niobate modulators, leading to more stable operation. When the phase difference between the two arms is zero (or π), the input light is ideally directed exclusively to the cross (or bar) port. In the full 2×8 optical switch architecture, arbitrary optical routing among the 8 output channels is achieved via rapid, independent, and simultaneous control of the AM units through the FC bonding platform, which would otherwise be prohibitively difficult to interface using probes.

We first characterized the switching voltage (also known as half-wave voltage $V_\pi$) of individual AM units by applying a low-frequency triangular modulation signal (Fig. 3c, top). The resulting optical transmission indicates a low $V_\pi$ value of 2.2 V to switch the modulator between cross and bar states (Fig. 3c, bottom). This yields a voltage-length product ($V_\pi \cdot L$) of 2.4 V· cm, which aligns closely with prior results on the TFLN platform[9].

We subsequently evaluated the optical performance of the full 2×8 switch. By optimizing the electrical voltages applied to the 7 AM units stage by stage, light launched into the "input" port can be successively routed exclusively to output ports "O1" - "O8". The corresponding transmission spectra of the intended output port and its neighboring port (excluding fiber-to-chip coupling loss of ~5 dB/facet) under these 8 switching states are shown in Fig. 3d. Notably, all 8 output channels show low on-chip insertion loss of approximately 3 dB and low inter-channel optical crosstalk < -20 dB over a broad wavelength range from 1540 nm to 1595 nm, indicating excellent uniformity and yield of our FC-packaged TFLN platform.

Finally, we showcase the high-speed switching capability of our system by applying a fast square-wave signal (Fig. 3e, top) to AM 3-1 to switch the output signal between "O1" and "O2" ports. The output optical transmission measured at "O2" port (Fig. 3e, bottom) closely aligns with the input waveform, with measured rise and fall times of 33 ps and 34 ps, respectively. The switching time is currently limited by the 13 GHz bandwidth of the oscilloscope used in this experiment, and could be further reduced to ~12 ps considering the broad bandwidth of our EO switches at the package level. Nevertheless, our TFLN EO switches have already achieved significantly higher speed than conventional silicon-based EO switches[44].

### EO comb-based photonic transmitter

The second photonic circuit demonstration is a wavelength-selective optical transmitter that leverages an on-chip EO comb to generate a multi-wavelength light source, a microring filter to select one comb line as optical carrier, and an AM for high-speed data transmission. Such a system could find applications in wavelength-division multiplexed (WDM) optical communication and parallel photonic computing systems[45-47].

As shown in the chip schematic in Fig. 4a, the photonic circuit starts with a double-pass PM optimized for maximum EO response at 50 GHz and a first AM (AM 1), together forming a flat-top EO frequency comb source. The comb output is filtered by a TO tunable add-drop microring resonator to selectively extract a single comb line from its drop port. The microring filter features a free spectral range (FSR) of approximately 280 GHz. The filtered light serves as the optical carrier, which is subsequently modulated by a second amplitude modulator (AM 2) for communication-related applications. In the future, the number of wavelength channels can be further scaled up to enable a high-capacity EO comb-based WDM system.

We first characterized the performance of the EO comb source at the monitor port "O1" using the experimental setup depicted in Fig. 4a (see details in Supplementary data). When driving only the PM using a 50 GHz RF signal at an on-chip RF power of 28 dBm via the PCBA, the device generated 13 comb lines with a flatness of 15 dB, as shown in Fig. 4b. The comb spectrum follows a typical Bessel shape with an estimated RF $V_\pi$ of ~3.9 V at 50 GHz, which also aligns well with the RF $V_\pi$ of the AM in the previous section, extracted to be ~3.7 V at 50 GHz from the DC $V_\pi$ and the EO S21 data (light gray curve in Fig. 2f). To further achieve a spectrally flat EO comb, which is essential for channel uniformity, transmission quality, and overall system capacity in a practical WDM system, we simultaneously drive both the PM and the AM 1 using RF signals split from the same RF source. With on-chip RF powers of 24.5 dBm applied to the PM and 9 dBm to AM 1, the device yielded 7 comb lines spaced at 50 GHz with an excellent flatness within 2.8 dB, as shown in the top panel of Fig. 4c.

We subsequently demonstrated arbitrary filtering of the generated flat-top EO comb lines by progressively increasing the heater power of the microring filter (Fig. 4c(ii)–(vi)). The out-of-band rejection ratios (OBRR) in all measured channels exceeded 20 dB, indicating excellent channel purity and negligible inter-channel crosstalk. The filter wavelength shift shows a clear linear dependence on the heating power, with a fitted TO tuning efficiency of 6.7 pm/mW (Fig. 4d, light red squares), consistent with our previous work[40]. We also characterized the thermal crosstalk from an adjacent TO heater on AM 2, which is located 700 μm away from the microring and represents the smallest distance between heaters on our current chip. As the blue triangles in Fig. 4(d) show, the measured crosstalk tuning efficiency is 0.36 pm/mW, which is approximately 5% of the TO tuning efficiency of the ring's own heater.

Finally, we provide a proof-of-concept validation of the communication capability of our transmitter system by applying on-off keying (OOK) signals to AM 2, using the left 1$^{st}$ sideband of the EO comb as the optical carrier. The measured eye diagrams in Fig. 4e show clear open-eye data transmission at various data rates, i.e., 5 Gbaud, 10 Gbaud, and 20 Gbaud, currently limited by the 13 GHz bandwidth of the oscilloscope. Considering the current comb spacing of 50 GHz and modulation bandwidth > 30 GHz, our chip-scale system in principle supports fast data communications of at least 40 Gbaud per channel (assuming 90% spectral efficiency). Notably, the experiments involved the simultaneous delivery of three high-speed EO signals (for AM 1, PM, and AM 2) and three DC/TO signals (for AM 1 bias, the microring filter, and AM 2 bias), further validating the effectiveness of the FC bonding platform for scalable and high-performance photonic integration.

## CONCLUSIONS

We have developed a multi-channel high-speed FC packaging platform capable of electrical interfacing with large-scale TFLN photonic circuits through up to 13 high-speed and 32 low-speed electrical channels. The platform features low FC bonding loss (average < 1 dB) and low inter-channel crosstalk (< -22 dB) across a broad frequency range from DC to 50 GHz. The two proof-of-concept TFLN photonic circuit demonstrations, i.e., the 2×8 optical switch and the EO comb-based photonic transmitter, which are otherwise difficult to interface using probes, both validate the excellent high-speed performance and channel scalability of our platform.

In the future, even broader EO bandwidth exceeding 100 GHz could be achieved by the use of lower-loss PCB material and TFLN-on-quartz substrates to reduce the microwave loss, as well as further optimization of the structural design and fabrication process of the chip. Currently, the number of channels is limited by the bonding pad size (250 μm, constrained by our initially adopted stencil-based solder-bump process). By adopting automated machine-based solder-ball placement and optimizing the material of the FC bonds, the pad size can be further reduced to accommodate a greater number of channels. The temperature profile of the FC bonding process can also be optimized to improve bonding yield. Additionally, co-packaging custom-designed ASICs with the PCBA could reduce the PCB trace lengths, which are currently necessary for connection with the bulky external cable connectors, leading to even broader EO bandwidths and higher channel density simultaneously. We note that when high-power ASIC chips are also packaged into the FC platform, effective thermal management will be essential. A manifold microchannel heat sink can be adopted for the ASIC to maximize heat dissipation efficiency, while parallel microchannel heat sinks for the photonic chips can provide temperature uniformity and stability, thereby mitigating thermal crosstalk[48]. Collectively, these advancements could ultimately enable ultra-dense, high-capacity interconnections between electronic and photonic integrated circuits with excellent stability and reliability. The demonstrated FC-packaging platform provides a viable solution for advanced applications in high-capacity WDM optical communication networks, photonic neuromorphic computing, and microwave photonics systems.

## METHODS

Detailed materials and methods are available in the Supplementary data.

## AUTHORS CONTRIBUTIONS

The project was conceived and coordinated by C.W. and Z.Z.W.; Z.Z.W., Y.K.C., Y.F.W., Z.X.W., Y.S.Z. and M.D.H. designed the optical and electrical components of the chip; Z.Z.W., Y.F.W. and H.K.F. performed the chip fabrication; Z.Z.W., Y.L.C., and Y.X.L. performed the FC bonding process; Z.Z.W., X.Z.X., Z.Y.L., Z.Y.X., and K.Z. performed the experimental characterization of the devices; C.W. supervised the project. All authors participated in discussions and analysis of the results.

## ACKNOWLEDGEMENT

We acknowledge SJTU-Pinghu Institute of Intelligent Optoelectronics for the FC bonding and packaging services.

## FUNDING

This work was supported by the National Natural Science Foundation of China (62525507); Research Grants Council, University Grants Committee (C1002-22Y, STG3/E-104/25-N, JRFS2526-1S01, CityU 11204824, N_CityU141/23); Croucher Foundation (9509005); City University of Hong Kong (9610682), Guangdong and Hong Kong Universities '1+1+1' Joint Research Collaboration Scheme (9249005, 9679005).

## DISCLOSURES

The authors declare no conflicts of interest.

## DATA AVAILABILITY

Data underlying the results presented in this paper are not publicly available at this time but may be obtained from the authors upon reasonable request.

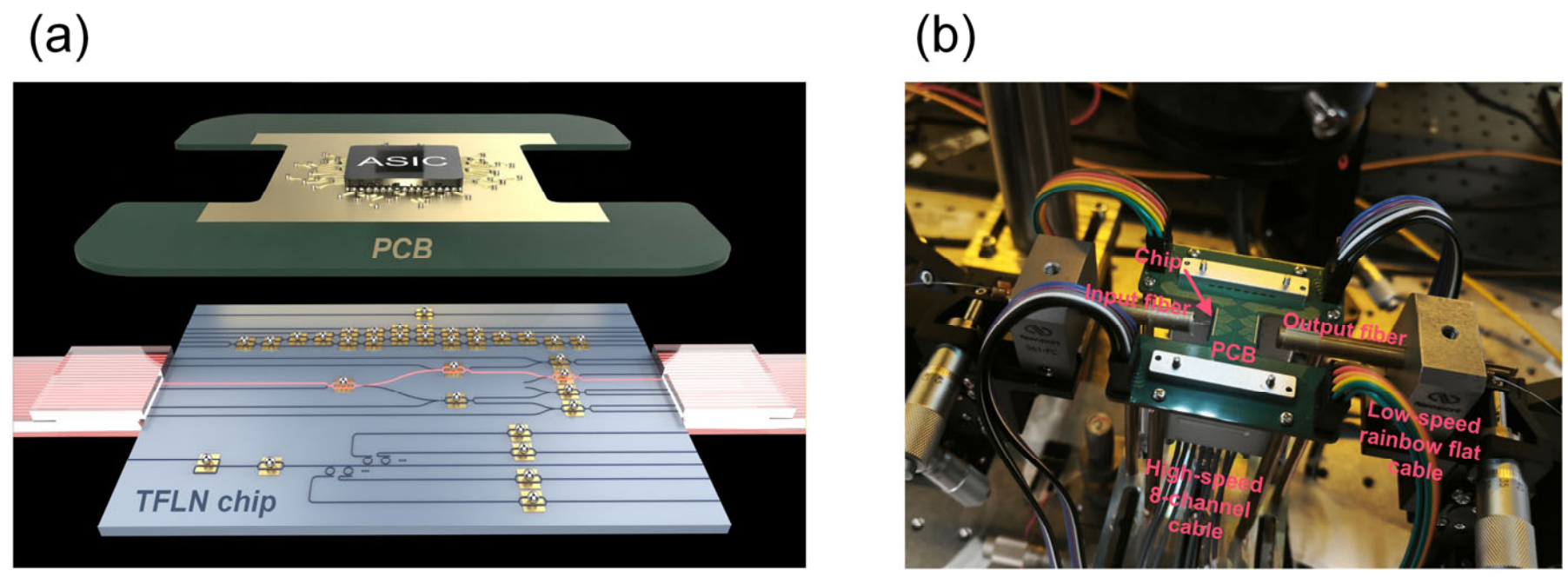

**Figure 1.** (a) Conceptual diagram of the envisioned FC packaging platform for interfacing TFLN photonic chips. (b) Picture of the FC packaged TFLN photonic chip under test in this work.

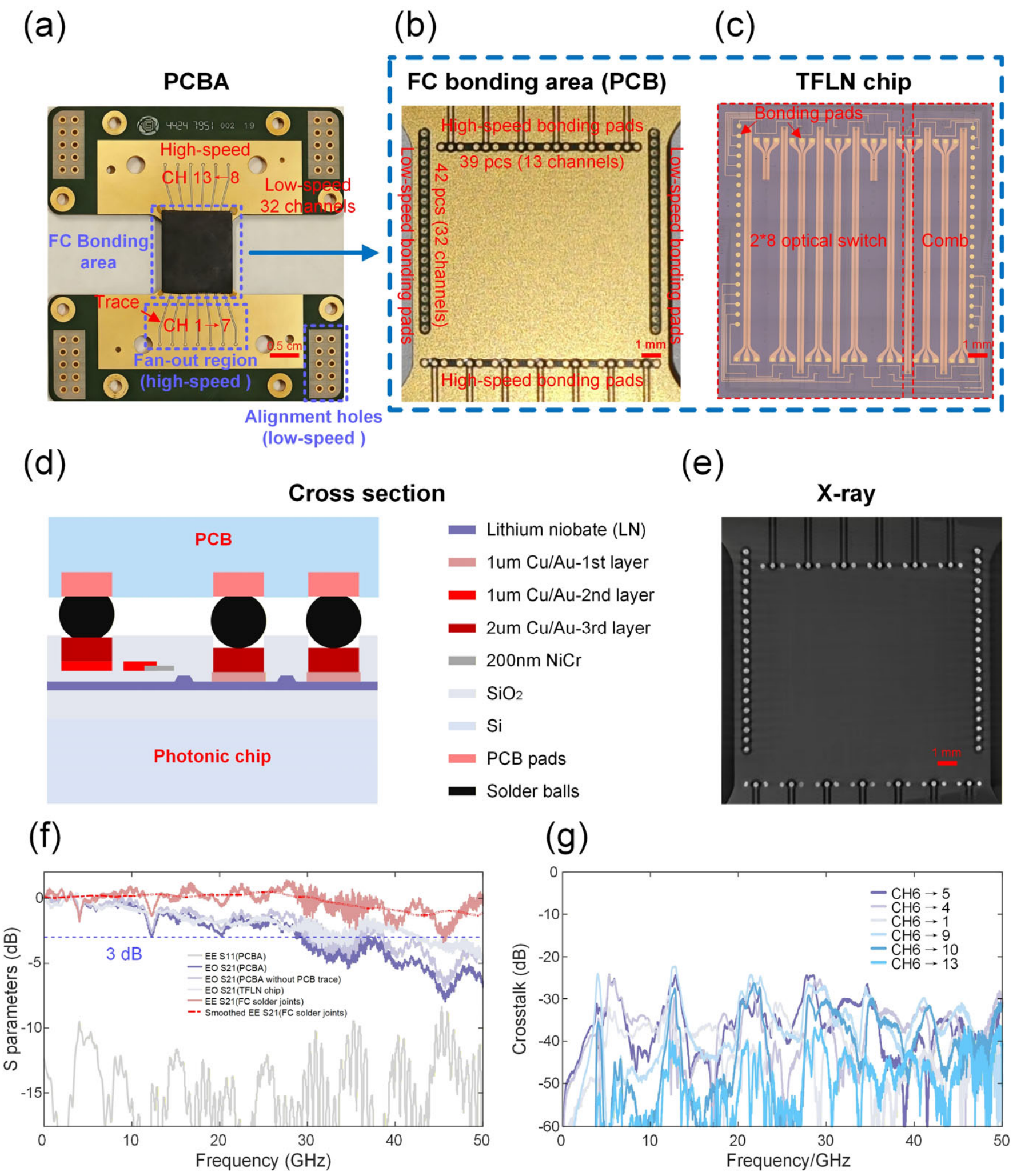

**Figure 2.** (a) Fabricated PCBA module after FC packaging of the integrated TFLN chip. (b) PCB bonding area before FC bonding. (c) Fabricated integrated TFLN chip before FC bonding. (d) Cross-sectional schematic of the bonding regions. (e) X-ray image of the FC bonding area on the PCBA module. (f) Measured EO S21 parameters of an amplitude modulator before (light gray) and after FC bonding (light and dark blue), along with extracted EE S21 parameters of the FC bond (raw data: light red curve, smoothed data: red dashed curve) and EE S11 parameter of the PCBA (gray). (g) Measured electrical crosstalk between different channels on the packaged PCBA.

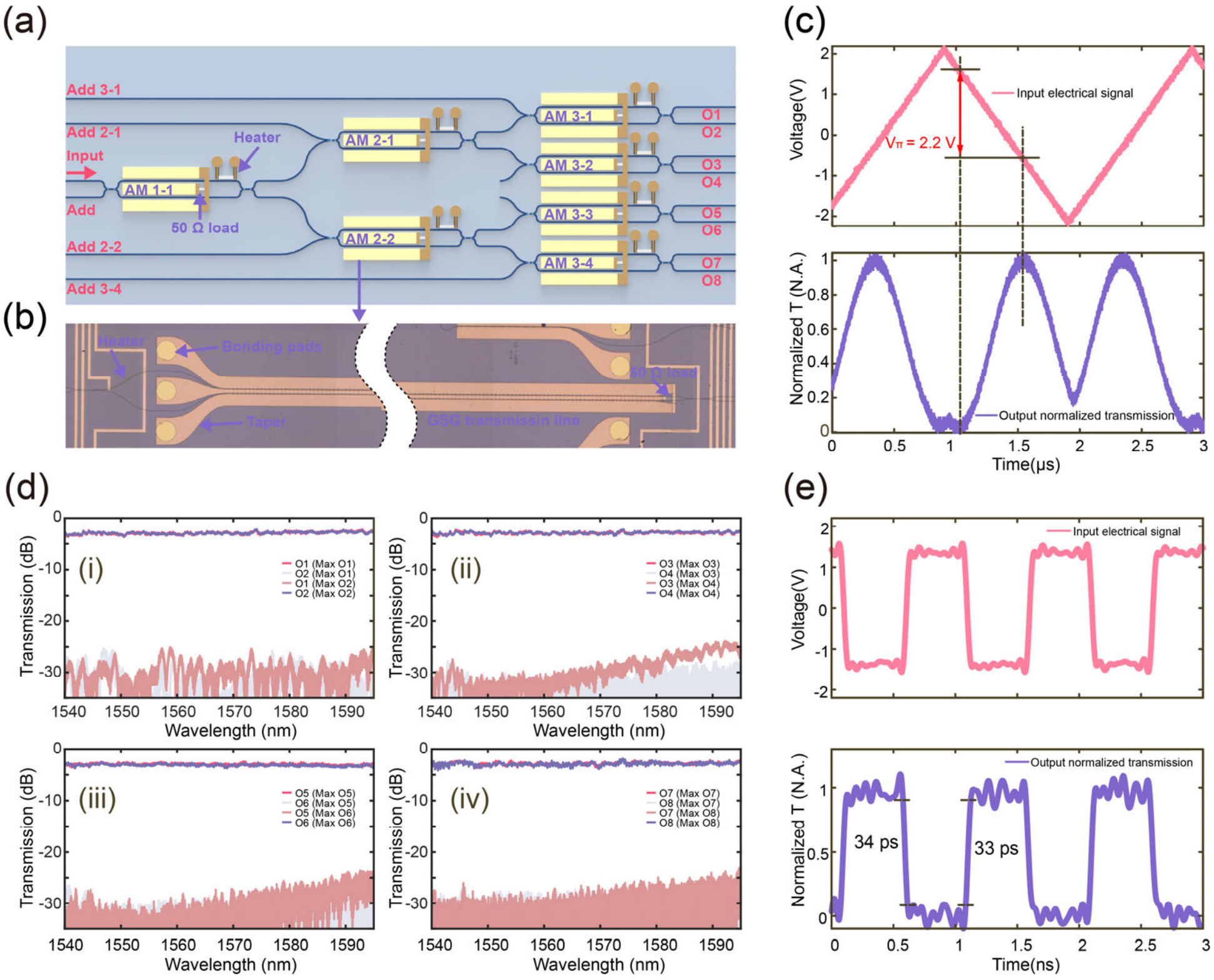

**Figure 3.** (a) Schematic diagram of the 2×8 optical switch. (b) Microscope image of one fabricated AM unit of the switch. (c) Measured normalized optical transmission of a representative AM 3-1 (bottom) under a triangular-waveform voltage input (top), indicating a low-frequency $V_\pi$ of 2.2 V. (d) Measured normalized transmission spectra of the 8 output channels at different switching states. (e) Measured time-domain responses of the switch. The red and blue curves respectively represent the applied square-wave electrical signals and the normalized output optical signals.

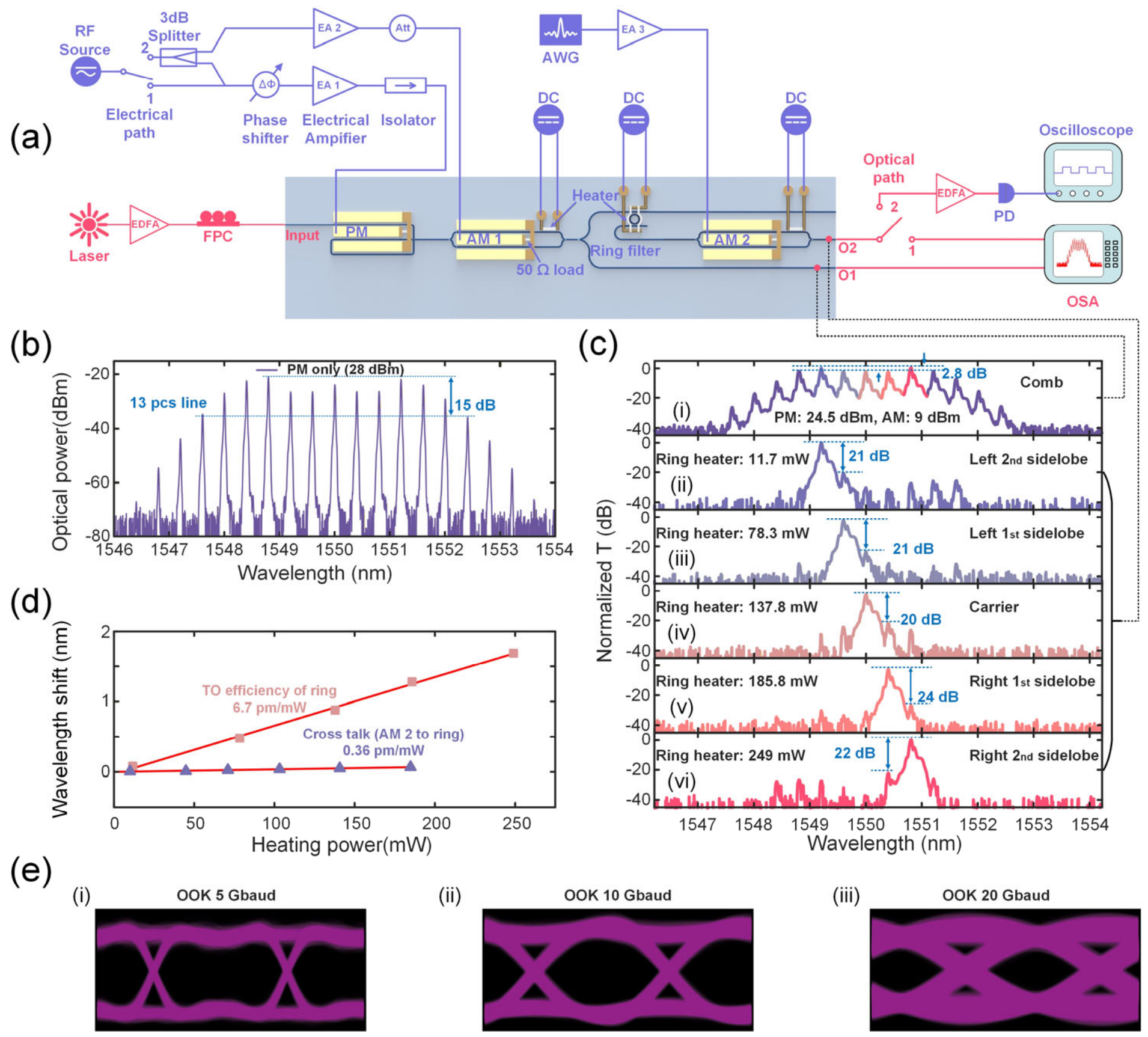

**Figure 4.** (a) Schematic of the experimental setup for the EO comb-based transmitter. (b) EO comb spectrum using PM only, at an on-chip RF power of 28 dBm. (c) (i) Flat-top EO comb spectrum when driving both PM and AM with on-chip powers of 24.5 dBm and 9 dBm, respectively. (ii)-(vi) Optical spectra after filtering different comb lines using a microring filter. (d) Filter wavelength shift vs. heating power of the microring (light red squares), in comparison with the crosstalk tuning efficiency of the microring from the TO heater on AM 2 (blue triangles, ~700 µm away). (e) Eye diagrams at various data rates using the first-order left sideband as the optical carrier.